\documentclass[sigconf]{acmart}
\AtBeginDocument{%
  }

\usepackage{color, soul}
\usepackage{algorithm}
\usepackage{algorithmic}
\usepackage{amsmath}
\begin{document}

\title{
RestAware: Non-Invasive Sleep Monitoring Using FMCW Radar and AI-Generated Summaries}


\author{Agniva Banerjee}
\email{agniva24@iiserb.ac.in}
\affiliation{%
  \institution{Indian Institute of Science Education and Research, Bhopal}
  \country{India}
  \postcode{462066}
}

\author{Bhanu Partap Paregi}
\email{bhanu24@iiserb.ac.in}
\affiliation{%
  \institution{Indian Institute of Science Education and Research, Bhopal}
  \country{India}
}

\author{Haroon R. Lone}
\orcid{0000-0002-1245-2974}
\email{haroon@iiserb.ac.in}
\affiliation{%
  \institution{Indian Institute of Science Education and Research, Bhopal}
  \country{India}
}

\ccsdesc[500]{Hardware~Sensor devices and platforms}
\ccsdesc[500]{Computing methodologies~Machine learning}
\ccsdesc[500]{Computing methodologies~Natural language generation}
\ccsdesc[500]{Applied computing~Health informatics}
 
\keywords{Sleep monitoring, FMCW radar, Language models, Sleep summary}
  
\begin{abstract}

Monitoring sleep posture and behavior is critical for diagnosing sleep disorders and improving overall sleep quality. However, traditional approaches—such as wearable devices, cameras, and pressure sensors—often compromise user comfort, fail under obstructions like blankets, and raise privacy concerns. To overcome these limitations, we present \textit{RestAware}, a non-invasive, contactless sleep monitoring system based on a 24GHz frequency-modulated continuous wave (FMCW) radar. Our system is evaluated on 25 participants across eight common sleep postures, achieving 92\% classification accuracy and an F1-score of 0.91 using a K-Nearest Neighbors (KNN) classifier. In addition, we integrate instruction-tuned large language models (LLMs)—Mistral, Llama, and Falcon—to generate personalized, human-readable sleep summaries from radar-derived posture data. This low-cost (\$ 35), privacy-preserving solution offers a practical alternative for real-time deployment in smart homes and clinical environments. 

\end{abstract}

\maketitle

\begin{figure}[htbp]
\centerline{\includegraphics[width=1\columnwidth]{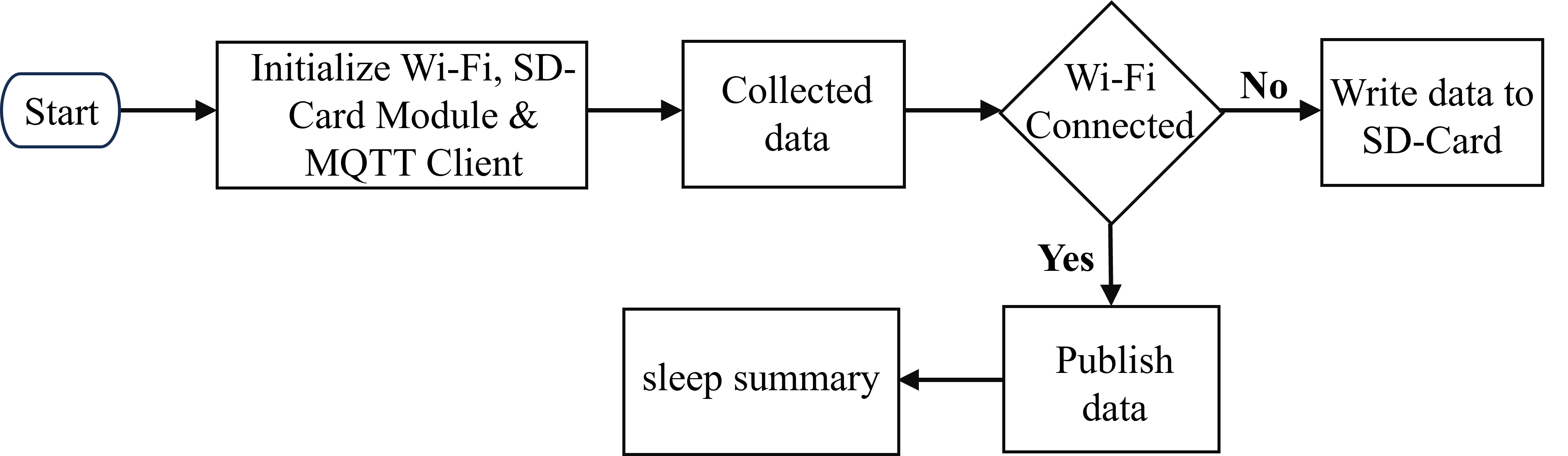}}
\caption{Process diagram of \textit{RestAware} system.}
\Description{Complete Process Diagram of System workflow.}
\label{sleep}
\end{figure}

\section{Introduction}

Sleep is essential to human health, with poor sleep linked to cardiovascular disease, depression, and impaired cognitive performance~\cite{jayarajah2015candy, karlgren2022self}. Accurate and non-disruptive sleep monitoring is therefore vital for diagnosing sleep disorders and promoting long-term well-being~\cite{perez2020future}. Traditional methods rely heavily on wearable sensors and clinical setups, which can interfere with natural sleep patterns, reduce user comfort, and hinder long-term adoption~\cite{arnal2020dreem, de2024state}.

Recent advances in contactless sensing technologies offer promising alternatives~\cite{nouman2021recent}. Systems based on radar~\cite{yao2025mm2sleep}, infrared~\cite{deng2018design}, and video~\cite{wang2024video} can unobtrusively capture vital signs such as respiration rate~\cite{bujan2023clinical}, heart rate variability~\cite{he2024novel}, body movements, and sleep posture~\cite{wang2022human}. However, these technologies still face practical barriers in home environments. Wearables remain uncomfortable for overnight use~\cite{arnal2020dreem}, radar systems often demand complex manual calibration and become costly when scaled to multiple devices~\cite{liu2023posmonitor}, infrared sensors are sensitive to ambient lighting and temperature~\cite{deng2018design}, and video-based solutions raise significant privacy concerns and depend on controlled lighting~\cite{wang2024video}.

To address these limitations, we introduce \textit{RestAware:} a cost-effective, contactless sleep monitoring system built around a 24 GHz FMCW human static presence radar. Operating in the 24–24.25 GHz ISM band, our system captures fine-grained Doppler frequency shifts and phase variations to detect both respiration patterns and posture changes~\cite{hong2019microwave}. Importantly, it eliminates the need for manual calibration and maintains performance across multiple sleep positions.

We evaluate \textit{RestAware} in a controlled study involving 25 male participants, each assuming eight distinct sleep postures in 10-minute intervals. These postures were selected to represent six common transition types in real-world sleep behavior (e.g., supine to side, prone to supine)~\cite{yao2025mm2sleep}, ensuring robust coverage of typical movements during sleep.

A key innovation in our system is its integration with instruction-tuned LLMs—including Mistral~\cite{jiang2023mistral7b}, Llama~\cite{touvron2023llama}, and Falcon~\cite{almazrouei2023falcon}—to generate human-readable sleep summaries from radar data. These models translate raw sensor inputs into personalized narratives describing posture transitions and sleep behavior, thereby enhancing user interpretability and engagement. We further benchmark their outputs against GPT-4 to assess coherence and readability.

\begin{table*}[!t]
\caption{Systematic Comparison of Radar-Based Sleep Posture classification Systems}
\label{tab:radar_systems}
\centering
\renewcommand{\arraystretch}{1.3}
\begin{tabular}{p{2.5cm} p{2.2cm} p{2.5cm} p{1.2cm} p{5cm} p{2.2cm}}

\toprule
\textbf{Study} & \textbf{Radar Technology} & \textbf{Model/Algorithm} & \textbf{Accuracy} & \textbf{Key Features} & \textbf{Classification} \\
\midrule
mm2Sleep~\cite{yao2025mm2sleep} & FMCW radar & Multiple Machine Learning (ML) models & 92.3\% & Works with multiple sleepers; robust calibration & Dual-person sleep posture classification \\
MiSleep~\cite{adhikari2024misleep} & Millimeter-wave radar & Deep Learning (DL) augmented & 91.2\% & Privacy-preserving; works through blankets; continuous monitoring & Sleep posture detection \\
PosMonitor~\cite{liu2023posmonitor} & mmWave radar & CNN & 94.7\% & Fine-grained posture detection; real-time processing & Sleep posture classification \\
BodyCompass~\cite{yue2020bodycompass} & Low-power wireless signal & CNN with RF reflection model & 94\% & Wall-mounted; works in diverse layouts & Sleep posture monitoring \\
Dual UWB~\cite{hong2019microwave} & Dual ultra-wideband radar & ML-based & -- & Contactless; radar-based detection & Sleep posture classification \\
SleepSense~\cite{lin2016sleepsense} & 5.8 GHz Doppler radar array & Random Forest~\cite{fan2019light} & 90.2\% & Multi-point sensing; robust in various environments & Sleep posture classification \\
SleepPoseNet~\cite{piriyajitakonkij2020sleepposenet} & UWB radar & Multi-view learning & 73.7\% & Recognizes postural transitions using UWB radar& Sleep postural transition classification \\
IoT-PSG~\cite{lin2017iot} & Multiple sensor fusion & Neural Networks & 87.5\% & Wireless PSG; comprehensive sleep monitoring & Sleep posture and disorder detection \\
\textbf{\textit{RestAware}} & 24 GHz FMCW  & 26 ML models and LLMs & 92\% & Non-invasive; no calibration; cost-effective (\textbf{\$35.10}) & Sleep posture classification and summary generation \\
\bottomrule
\end{tabular}
\end{table*}

Our system achieves a posture classification accuracy of 92\% and an F1-score of 0.91 using a KNN classifier~\cite{zhang2017learning}, without requiring manual calibration. The entire setup—including radar and ESP32-based processing—costs approximately $ \$35.10$, making it a viable option for affordable, privacy-preserving in-home sleep monitoring. An overview of the system workflow is presented in Figure~\ref{sleep}.

The key contributions of this work are summarized below:
\begin{itemize}
    \item \textbf{Radar-Based, Calibration-Free Sleep Monitoring:} We present a non-invasive 24GHz FMCW radar system that classifies eight sleep postures with 92\% accuracy and 0.91 F1-score, \textit{all without manual calibration}.

 \item \textbf{Narrative Sleep Summarization with LLMs:} \textit{RestAware} uses instruction-tuned LLMs to generate narrative sleep reports from sensor data and evaluates their effectiveness compared to GPT-4.

 \item \textbf{Low-Cost, Privacy-Preserving Design:} The full system is implemented at a low cost ($\$ 35.10$) and does not rely on cameras or wearables, supporting unobtrusive and privacy-conscious monitoring in home settings.
\end{itemize}

By combining low-cost radar sensing with explainable AI-generated feedback, \textit{RestAware} offers a novel alternative to both wearable and camera-based approaches. This work directly contributes to the ISWC community by advancing the design of user-centered, non-wearable systems for continuous and interpretable human state monitoring.

\section{ Related work}

The growing interest in contactless sleep monitoring has positioned mmWave radar as a leading technology for accurate and privacy-preserving detection of vital signs~\cite{wang2022human}. Prior studies have explored a wide range of non-contact modalities, including radar, infrared, camera-based, and acoustic systems, to track respiration, heart rate, and posture variations during sleep~\cite{li2025novel, yao2025mm2sleep, liu2023fmcw, islam2022sleep}. For example, Lin \textit{et al.} developed an IoT-enabled framework as an alternative to polysomnography for sleep apnea diagnosis~\cite{lin2017iot}. In contrast, Yue \textit{et al.} introduced BodyCompass, an RF-based system enabling unobtrusive posture monitoring~\cite{yue2020bodycompass}. Similarly, Han \textit{et al.} proposed EarSleep, an in-ear acoustic sensing platform leveraging deep learning for fine-grained sleep stage detection~\cite{han2024earsleep}. Despite their effectiveness, each of these systems presents certain limitations: wearable devices like EarSleep may compromise user comfort~\cite{han2024earsleep}; camera-based approaches raise privacy concerns~\cite{yue2020bodycompass}; and infrared systems often suffer from limited operational range and high sensitivity to ambient conditions. In contrast, mmWave radar-based solutions, such as MiSleep by Adhikari \textit{et al.}~\cite{adhikari2024misleep} and mm2Sleep by Yao \textit{et al.}~\cite{yao2025mm2sleep}, achieve high-precision posture classification even under occlusions such as blankets and in multi-user scenarios all without the need for manual calibration.

However, these radar systems often remain cost-prohibitive for widespread home deployment (e.g., some setups exceed \$237~\cite{yao2025mm2sleep}). To address this accessibility barrier, \textit{RestAware} leverages a low-cost 24GHz FMCW radar sensor to enable robust, calibration-free detection of delicate physiological patterns, thus enhancing practicality for in-home sleep monitoring applications. Table ~\ref{tab:radar_systems} provides a comparison of \textit{RestAware} and other existing radar-based sleep monitoring systems. 

\begin{figure}[t]
\centerline{\includegraphics[width=0.8\columnwidth]{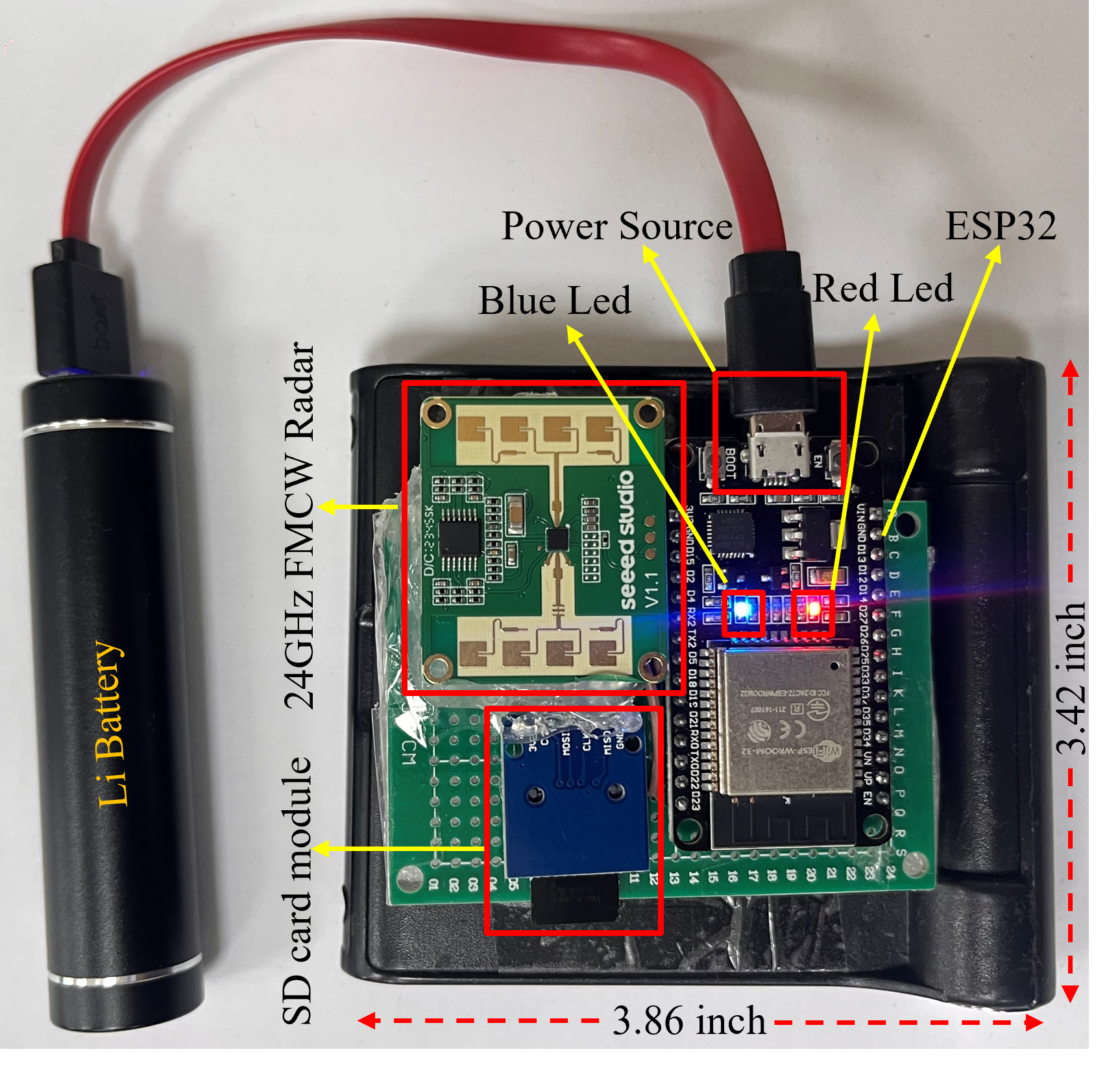}}
\caption{Hardware setup of \textit{RestAware} system.}
\Description{Hardware setup of \textit{RestAware} system.}
\label{PCB}
\end{figure} 

\section{ Methodology}\label{sec:Methd}
This section outlines the systematic approach used to develop the \textit{RestAware}. The process is divided into four key phases: hardware setup, data collection, data processing , and system workflow.

\subsection{Hardware Setup}\label{sec:Hardware}

The hardware setup (see Figure~\ref{PCB}) consists of a Seed Studio 24GHz FMCW radar sensor\footnote{\url{https://wiki.seeedstudio.com/Radar_MR24HPB1/}} interfaced with a 30-pin ESP32 microcontroller via a Universal Asynchronous Receiver/Transmitter (UART) interface for data communication. The ESP32 microcontroller features a dual-core 240 MHz processor and built-in Wi-Fi connectivity, serving as the system’s master controller. A mini micro-SD card module is connected to the ESP32 through the Serial Peripheral Interface protocol to facilitate local data storage. The entire system is powered by a 3.7 V lithium-ion battery, enabling portable and continuous operation throughout the night. The ESP32’s built-in LED provides visual feedback on system status and data logging activity.


Sleep data captured by the radar sensor is transmitted to the ESP32 over UART at a baud rate of 112,500. The ESP32 stores this data on the micro-SD card for offline processing when Wi-Fi connectivity is unavailable. The system’s low power consumption supports uninterrupted data collection overnight. When Wi-Fi connectivity is available, the ESP32 activates its wireless module and streams raw sleep data in real time to a remote server via the Message Queuing Telemetry Transport (MQTT) protocol. The server subsequently processes this data for inference and sleep summary generation.

\begin{figure}[htbp]
\centerline{\includegraphics[width=1\columnwidth]{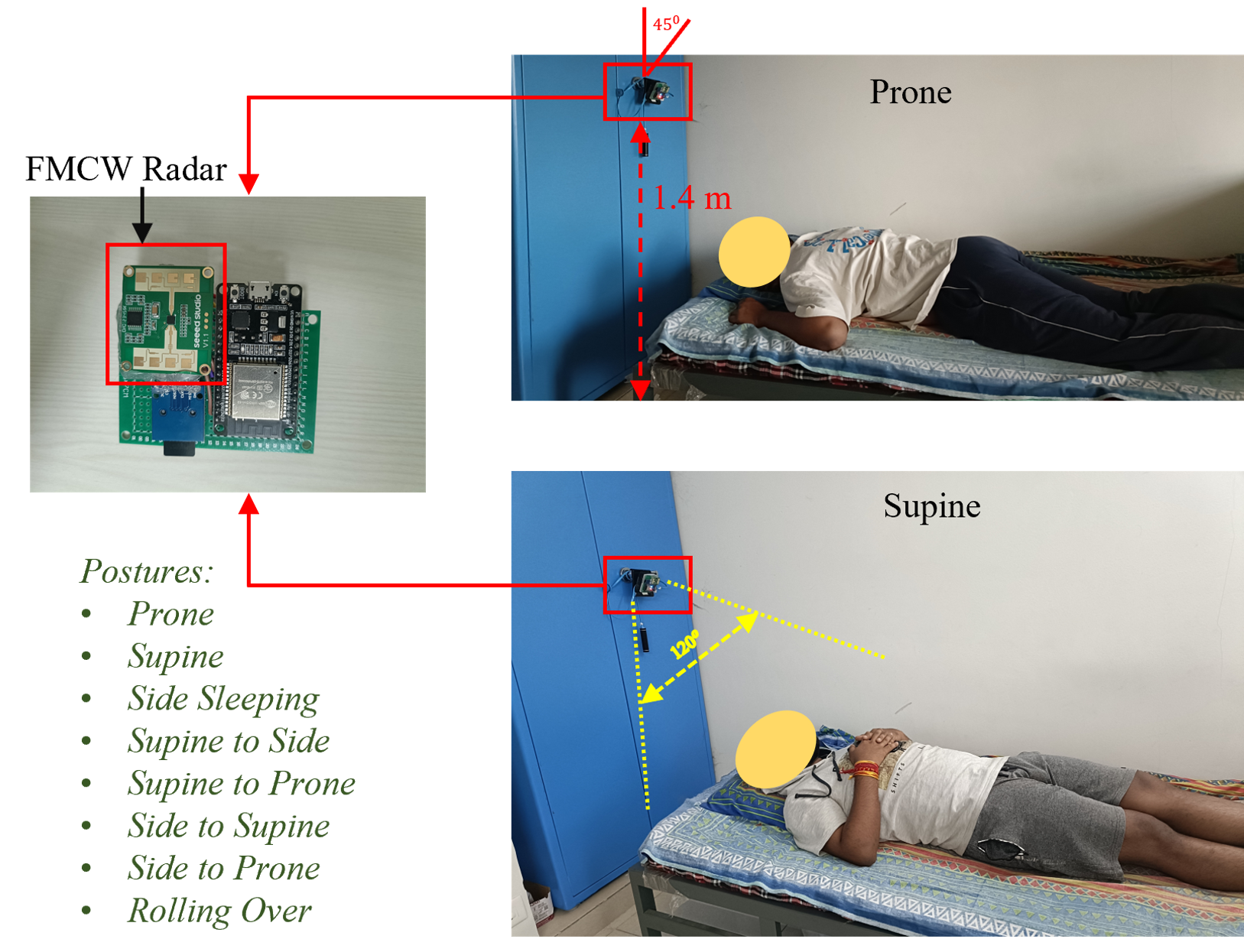}}
\caption{\textit{RestAware} placement during data collection.}
\label{Device}
\end{figure}

\begin{figure*}[htbp]
\centerline{\includegraphics[width=17cm, height = 7cm]{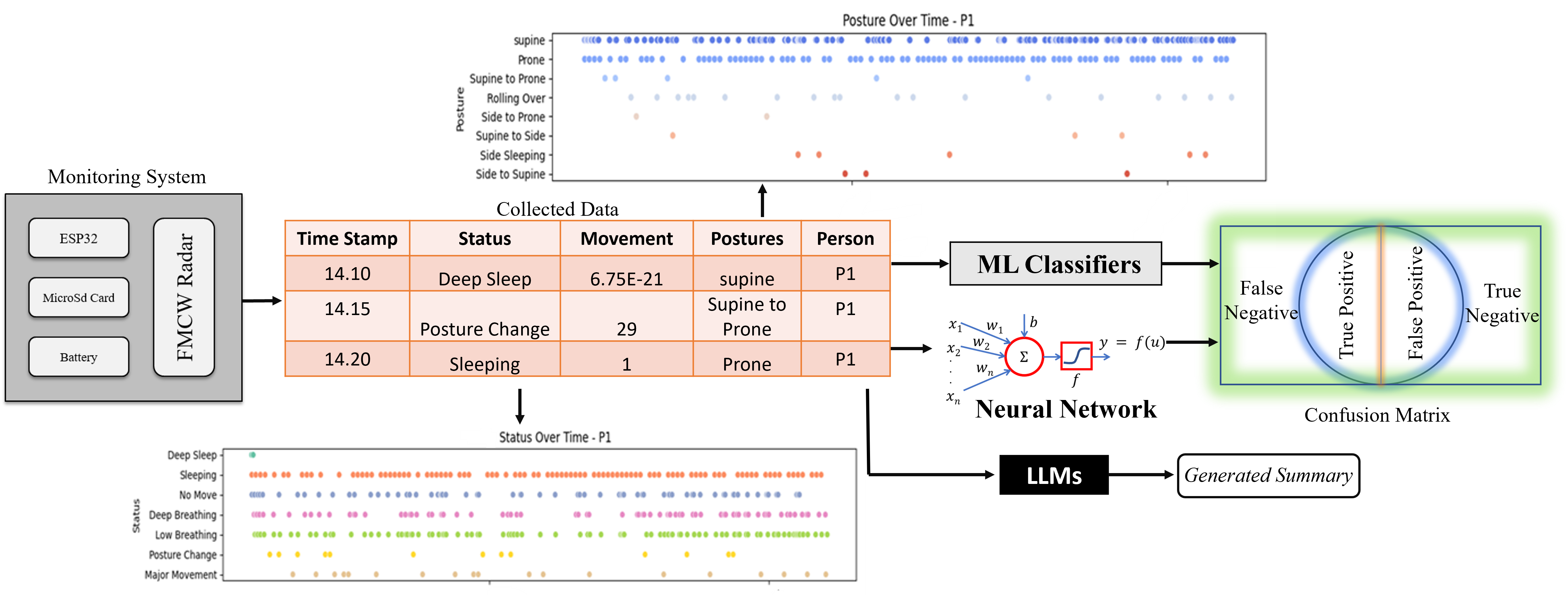}}
\caption{
End-to-end pipeline of the \textit{RestAware} system. The system captures raw posture and movement data via radar, performs sleep state classification using ML and DL models, and generates narrative summaries and visualizations with instruction-tuned LLMs to assess sleep quality, posture transitions, and behavioral patterns.}
\Description{
End-to-end pipeline for sleep posture analysis using FMCW radar. The system records posture and movement data, classifies sleep states using ML and LLM models, and generates visualizations and summaries to assess sleep quality and patterns.}
\label{Workflow}
\end{figure*}

\subsection{Data Collection}\label{sec:Data_collection}
During data collection, the \textit{RestAware} device (see Figure~\ref{Device}) was positioned at a height of 1.4 meters from the ground, inclined at an angle of $45^\circ$ with respect to the horizontal axis, and placed at a distance of 40 cm from the participant to ensure optimal radar coverage.

In accordance with the Ethics Review Board guidelines, data were collected from 25 male participants aged between 20 and 30 years, after obtaining informed consent. Each participant was instructed to adopt any of eight predefined sleep postures: \textit{prone}, \textit{supine}, \textit{side sleep}, \textit{supine to side}, \textit{supine to prone}, \textit{side to supine}, \textit{side to prone}, and \textit{roll-over}. Within a 10-minute session, each participant was free to transition among these postures, allowing the FMCW radar to capture dynamic motion (i.e., temporal changes in the radar scene) and presence data (i.e., occupancy status) in hexadecimal format. To ensure consistency and reduce noise, all sessions were conducted in a controlled indoor environment with stable ambient lighting and minimal external disturbances.

For ground truth annotation, video recordings were captured using a smartphone during each session. These recordings enabled manual labeling of sleep postures, ensuring accurate classification labels. The labeled video data were synchronized and aligned precisely with the radar-derived segments to create a high-quality dataset for training and validation. This alignment allowed robust assessment of \textit{RestAware's} posture classification performance.

The final dataset includes rich temporal information covering posture transitions, body movement intensities, and stillness periods, forming the basis for training learning-based models for sleep posture analysis.


\subsection{Data Processing}\label{sec: Data_processing}
\textit{RestAware's} data processing pipeline begins by decoding raw hexadecimal data packets generated by the FMCW radar sensor, which captures motion signatures associated with human movement, respiration, and posture changes. Each packet is parsed using a custom decoding function to extract key parameters, including packet type and movement amplitude. The raw data are then converted into a time series format, where each frame corresponds to a discrete time window of radar readings. These frames undergo cleaning and normalization procedures to remove noise and inconsistencies, ensuring that extracted features remain comparable across participants and recording sessions.

From the cleaned signals, a set of features is derived, including movement amplitude, posture stability, and temporal variation metrics. These features are further aggregated to characterize participant sleep behavior over time. The data are segmented based on posture transition boundaries and annotated according to the eight predefined sleep postures. In total, the dataset comprises 9,125 data points collected from 25 participants, averaging approximately 365 samples per participant over a 10-minute session. Prior to model training, the extracted features are standardized to ensure consistent scaling. The dataset is then split into training and validation sets using a 70:30 ratio.









\vspace{-0.2cm}
\subsection{System Workflow and Sleep Summary Generation}\label{sec: worlflow}

The \textit{RestAware} system (see Figure~\ref{Workflow}) initiates with continuous acquisition of raw sleep data from the user via a 24GHz FMCW radar sensor. These raw signals are decoded into time series frames and subsequently processed through a pipeline that includes noise reduction and signal normalization to ensure high-quality and consistent inputs across sessions. From the preprocessed signals, salient features such as sleep posture labels, movement amplitude, and presence status are extracted. These features form the input for various ML models, including neural networks, to classify sleep postures.

The neural network employed is a Multi-Layer Perceptron (MLP) with two fully connected layers of 64 and 32 units respectively, using ReLU activations and a dropout layer with a 0.3 rate to reduce overfitting. To identify the most effective posture classification ML model, we employed the \textbf{Lazypredict} framework~\cite{lazypredict}, which benchmarks 26 standard ML classifiers\footnote{Classifiers: KNeighborsClassifier, LGBMClassifier, DecisionTreeClassifier, NearestCentroid, GaussianNB, PassiveAggressiveClassifier, Perceptron, ExtraTreeClassifier, LabelPropagation, BaggingClassifier, LabelSpreading, XGBClassifier, RandomForestClassifier, SVC, AdaBoostClassifier, CalibratedClassifierCV, DummyClassifier, BernoulliNB, LogisticRegression, LinearDiscriminantAnalysis, LinearSVC, RidgeClassifierCV, RidgeClassifier, QuadraticDiscriminantAnalysis, SGDClassifier, NuSVC.}. These models were trained and validated on radar-derived features to determine their ability to classify the eight predefined sleep postures accurately. 

\begin{algorithm}
\caption{Sleep Summary Generation Using LLMs}
\label{alg:sleep_summary}
\begin{algorithmic}[1]
\REQUIRE Sleep data of user $D_u$, \textit{LLM} and \textit{tokenizer}, number of summary sentences $N$  
\ENSURE Extractive \textit{summary}

\STATE Load \textit{LLM} and \textit{tokenizer}  
\STATE $prompt \gets \textit{Construct\_Prompt}(D_u)$  
\STATE $text\_output \gets \textit{LLM}(prompt)$  
\STATE $Sentences \gets$ Split $text\_output$ into sentences 
\STATE $Sentence\_Scores \gets$ empty dictionary  

\FORALL{$s \in Sentences$}  
    \STATE $score \gets \textit{Importance\_Score}(s, text\_output)$  
    \STATE $Sentence\_Scores[s] \gets score$  
\ENDFOR  

\STATE $Sorted\_Sentences \gets$ \textit{Sentences} sorted by $Sentence\_Scores$ in descending order  
\STATE $summary \gets$ first $N$ sentences from $Sorted\_Sentences$  

\RETURN $summary$  
\end{algorithmic}
\end{algorithm}

Following the classification phase, \textit{RestAware} transitions into its narrative synthesis module, which leverages instruction-tuned LLMs to generate user-specific summaries of sleep behavior. These summaries describe aspects such as total sleep duration, posture transitions, restlessness, stillness intervals, and disruptions. To support this, the system constructs temporally ordered sequences of sleep data and aggregates behavioral metrics, including the frequency of deep sleep events, transitions between postures, breathing/no-movement segments, and average movement intensity.

A structured \textit{prompt} is generated from this aggregated data ($D_u$) using a rule-based template that transforms numerical and temporal metrics into descriptive input for the LLM. This \textit{prompt} is then passed to instruction-tuned LLMs such as Mistral-7B-Instruct (mistralai/Mistral-7B-Instruct-v0.1), LLaMA-2-7B-chat (NousResearch /Llama-2-7b-chat-hf), and Falcon-7B-Instruct (tiiuae /falcon-7b-instruct). These models are chosen for their open-source accessibility and ability to generate structured, human-readable summaries. The models are executed in a GPU-enabled environment (NVIDIA Ada Generation 4000 with 20 GB VRAM), using appropriate tokenizers and weights.

The LLMs produce paragraph-length summaries describing the user’s sleep. To enhance readability and focus, an extractive summarization step follows, as detailed in Algorithm~\ref{alg:sleep_summary}. The generated text (\textit{text\_output}) is segmented into individual \textit{sentences}, each of which is scored for relevance using a TextRank-based method~\cite{mihalcea2004textrank}. The top 
$N$ most informative sentences are selected to construct the final \textit{summary}, thereby combining the generative strengths of LLMs with extractive techniques to produce concise and coherent outputs.

In addition to the open source LLM models, we also passed the same temporal sleep sequence data to GPT-4, a state-of-the-art proprietary LLM developed by OpenAI~\cite{openai2023gpt4}. GPT-4 is included in our evaluation due to its superior language generation capabilities, particularly in producing high-quality, contextually aware, and fluent narrative text. This allowed to establish a performance baseline for high end commercial LLMs and assess the relative effectiveness of open access models in comparison. GPT-4's responses are used to evaluate potential improvements in summary richness and coherence when utilizing a more advanced instruction-tuned LLMs. 

\subsubsection{Evaluation Metrics}
 
We use the following metrics to evaluate the classification performance of different classifiers.

\begin{itemize}
    \item \textbf{Accuracy:} It represents the proportion of correct posture and predictions out of all predictions. It assess the general effectiveness of the classification models.
    $$\textbf{Accuracy} = \frac{TP + TN}{TP + TN + FP + FN}$$
where, $TP$: True Positive, $TN$: True Negative, $FP$: False Positive, $FN$: False Negative.

    \item \textbf{F1-Score:} It combines precision and recall into a single metric by balancing them. It is instrumental when dealing with imbalanced data instances, ensuring the model performs well even when some postures are underrepresented.
    $$\textbf{F1-score} = 2 \times \frac{\text{Precision} \times \text{Recall}}{\text{Precision} + \text{Recall}}$$

Where,  $$\textbf{Precision} = \frac{TP}{TP + FP}~~~~~\& \quad \textbf{Recall} = \frac{TP}{TP + FN}$$
    \item \textbf{Receiver-operating characteristic curve (ROC):} It plots True Positive Rate (TPR) vs False Positive Rate (FPR). A higher ROC indicates better model performance, suggesting the model can better differentiate between sleep postures. \\
$$\textbf{TPR} = \frac{TP}{TP + FN}, \quad \textbf{FPR} = \frac{FP}{FP + TN}$$
\end{itemize}



\section{ Results}

\subsection{Sleep posture classification}
Posture classification is treated as distinct multi class classification tasks using features extracted from preprocessed radar time series data. The performance of three best performing ML classifiers and a NN are presented in Table~\ref{tab:posture_classification_results}. Of the ML classifiers, the KNN classifier~\cite{zhang2017learning} achieved the highest performance, with an accuracy of 92\% and an F1 score of 0.91, a low inference latency of 0.03 seconds per window. In comparison, the Light Gradient Boosting Machine (LGBM)~\cite{fan2019light} achieved an accuracy of 84\%, an F1 score of 0.81, and an inference time of 0.79 seconds. The decision tree classifier achieved an accuracy of 81\% and an F1 score of 0.79. However, the NN classifier achieved an accuracy of 78\% and an F1 score of 0.77 for posture classification, although with a higher inference latency compared to the ML classifiers. These results highlight the trade-offs between model complexity, inference time, and classification performance, emphasizing the effectiveness of traditional ML approach, particularly the KNN classifier, in achieving high accuracy with low computational overhead for this task.

\vspace{-0.2cm}
\begin{table}[ht]
\centering
\caption{Sleep posture classification results}
\label{tab:posture_classification_results}
\scriptsize
\resizebox{\linewidth}{!}{%
\begin{tabular}{lcccc}
\toprule
\textbf{Model} & \textbf{Accuracy} & \textbf{F1-Score} &  \textbf{ROC} &  \textbf{Time (s)} \\
\midrule
KNN     & \textbf{0.92} & \textbf{0.91} &  0.95 & 0.03 \\
LGBM          & 0.84 & 0.81 &  0.83 & 0.79 \\
DecisionTree  & 0.81 & 0.79 & 0.8 & 0.01 \\
Neural Network          & 0.78 & 0.77 & 0.79 & 2.11 \\
\midrule
\textbf{Overall\footnotemark} & 0.916 & 0.911 & 0.95 & --\\
\bottomrule
\end{tabular}%
}
    \end{table}
\footnotetext{Average over the 26 classifiers of the Lazypredict framework~\cite{lazypredict}}
\vspace{-0.5cm}
\subsection{Sleeping summary generation}

The process of sleep summary generation is a core component of the \textit{RestAware} system, designed to deliver personalized, coherent, and insightful reports based on the captured sleep data. These summaries aim to help users better understand their sleep patterns, thereby contributing to improved sleep management and overall well-being. Table~\ref{tab:llm_summaries} presents example summaries for participant P1, generated using four different language models: Mistral, LLaMA, Falcon, and GPT-4.

Among the four, the summary generated by Mistral stands out due to its empathetic tone and intuitive interpretation. It effectively highlights deep sleep and stillness as positive indicators, while also considering posture stability and low movement levels as signs of physical relaxation or mild discomfort—adding a human-centric perspective to the analysis. In contrast, GPT-4 offers a more clinical and structured assessment, whereas LLaMA stays closely aligned with the raw data. Although all summaries are informative, Mistral’s output is the most user-friendly and emotionally resonant, making it a strong candidate for delivering sleep insights in a relatable and reassuring manner.

\begin{table}[ht]
\centering
\caption{Sample Sleep Summaries Generated by Different LLMs for participant, P1.}
\begin{tabular}{@{}l p{0.7\linewidth}@{}}
\toprule
\textbf{LLM Model} & \textbf{Summary} \\
\midrule
Mistral & \textit{``P1 appears to have a very restful sleep with a lot of deep sleep, which is great for overall health and well-being. The number of prone posture instances suggests that P1 may have some back pain or discomfort, but this could also be a sign of relaxation. The lack of movement and low breathing events indicate that P1 was in a deep state of rest, which is important for physical recovery and mental rejuvenation. The `No Move' events also suggest that P1 was able to stay still and quiet throughout the night, which can help to reduce the risk of waking up during sleep. Overall, P1's sleep pattern appears to be very healthy and restful, with plenty of deep sleep and minimal disruptions.''}\\
\addlinespace[1.5ex]
LLaMA-2 & \textit{``Based on the data provided, this person had a relatively consistent sleep pattern throughout the night, with three instances of deep sleep. During these deep sleep periods, the person showed very little movement, indicating that they were in a state of deep rest. Additionally, there were 54 instances of ``Low Breathing'' and 62 instances of ``No Move'' which further suggests that the person was in a state of deep relaxation. Overall, this person's sleep pattern indicates that they were able to get a good night's rest, with a balance of deep sleep and relaxation.''} \\
\addlinespace[1.5ex]
Falcon & \textit{``The individual in question exhibits a predominantly deep sleep pattern throughout the night, with only minimal movement and a low amount of breathing events. This suggests a high level of rest and overall physiological well-being.''} \\
\addlinespace[1.5ex]
GPT-4 & \textit{``Based on the provided sleep posture data, P1 exhibits patterns of deep sleep primarily in the supine position, with occasional transitions to the prone posture. There are instances of major movements, including rolling over and posture changes, which may indicate restlessness. The presence of ``Low Breathing'' and ``No Move'' events suggests fluctuations in sleep stability. These variations, along with significant movement spikes, could be potential indicators of sleep disturbances, such as fragmented sleep or possible underlying sleep disorders.''} \\
\bottomrule
\end{tabular}
\label{tab:llm_summaries}
\end{table}

\begin{table}[h]
\centering
\caption{Comparison of LLMs generated sleep summaries}
\label{tab:summary_comparison_llms}
\scriptsize
\begin{tabular}{p{0.6cm}p{0.8cm}p{0.9cm}p{1cm}p{1.2cm}p{0.8cm}p{0.6cm}}
\hline
\textbf{Model} & \textbf{Clarity} & \textbf{Depth of Analysis} & \textbf{Diagnostic Value} & \textbf{Personalization} & \textbf{Balance} & \textbf{Overall} \\
\hline
Mistral & High & High & Moderate & High & Mostly Positive & Strong \\
LLaMA & High & Moderate & Low & Moderate & Positive-focused & Fair \\
Falcon & Moderate & Low & Very Low & Low & Overly Positive & Weak \\
GPT-4 & High & High & High & High & Balanced & Strongest \\
\hline
\end{tabular}
\end{table}

On comparing the generated summaries (see Table~\ref{tab:summary_comparison_llms}) across several key dimensions—including clarity, depth of analysis, diagnostic value, personalization, and balance—the Mistral model emerges as a highly favorable choice for deployment. It demonstrates strong overall performance, particularly excelling in clarity, analytical depth, and personalized tone. Moreover, as a publicly accessible model, Mistral presents a cost-effective alternative to proprietary models such as GPT-4. This aligns well with \textit{RestAware's} overarching goal of delivering interpretable and high-precision sleep insights without incurring high deployment costs.

\section{Discussion and limitations}
\textit{RestAware} achieves an overall accuracy of 92\%, F1-score of 0.91, and ROC score of 0.95. While these scores are slightly lower than those reported by high-end or multi-radar systems~\cite{liu2023posmonitor, yao2025mm2sleep}. However, it offers a unique trade-off between performance, cost, and usability. With a compact architecture costing under \$35.10, no camera, and fully contactless operation, it enables practical and privacy-preserving deployment for real-world home environments. The ability to transmit classified radar data via MQTT and generate meaningful summaries using instruction-tuned LLMs enhances system utility, allowing separation of lightweight edge computation from cloud based analysis. This hybrid sleep monitoring architecture balances low power hardware constraints with advanced AI capabilities.

Despite its strengths, the system presents a few limitations:

\begin{itemize}
    \item \textbf{LLM Deployment Constraints:} LLMs (e.g., Mistral, Llama, Falcon) used for sleep summary generation cannot run on the ESP32 microcontroller due to memory and compute limitations. Instead, they require offloading to a server via Wi-Fi.
    
   \item \textbf{Radar's Range Limitation:} The 24 GHz FMCW radar, with a detection range of 0.4–2.5 meters and a field of view of ±60° horizontal and ±30° vertical, is effective for monitoring a single person in a small room with a single bed. However, its limited range and coverage make it unsuitable for larger areas or halls with multiple persons, restricting its applicability in multi-occupant environments.
\end{itemize}

\section{ Conclusion}
In conclusion, the proposed radar-based sleep monitoring system presents a compelling, non-invasive, and contactless alternative to traditional sleep tracking methods, addressing key limitations associated with wearables, cameras, and high costs. Leveraging a 24GHz FMCW radar, the \textit{RestAware} system achieved 92\% accuracy and a 0.91 F1-score for sleep posture classification using a KNN classifier, underscoring its effectiveness. Additionally, the integration of instruction-tuned LLMs enables the generation of detailed and interpretable sleep summaries, offering valuable insights into sleep patterns, stability, and disturbances. The system's ability to operate across new users without human calibration, combined with its real-time performance and low hardware cost (approximately \$35.10), highlights its potential for scalable, multi-user applications in healthcare and smart home environments. Overall, this radar-based solution delivers a privacy-preserving and cost efficient approach, reducing expenses by up to 84\% compared to conventional methods while maintaining high performance and user convenience.


\bibliographystyle{ACM-Reference-Format}
\bibliography{sample-base}


\begin{thebibliography}{31}


\ifx \showCODEN    \undefined \def \showCODEN     #1{\unskip}     \fi
\ifx \showISBNx    \undefined \def \showISBNx     #1{\unskip}     \fi
\ifx \showISBNxiii \undefined \def \showISBNxiii  #1{\unskip}     \fi
\ifx \showISSN     \undefined \def \showISSN      #1{\unskip}     \fi
\ifx \showLCCN     \undefined \def \showLCCN      #1{\unskip}     \fi
\ifx \shownote     \undefined \def \shownote      #1{#1}          \fi
\ifx \showarticletitle \undefined \def \showarticletitle #1{#1}   \fi
\ifx \showURL      \undefined \def \showURL       {\relax}        \fi
\providecommand\bibfield[2]{#2}
\providecommand\bibinfo[2]{#2}
\providecommand\natexlab[1]{#1}
\providecommand\showeprint[2][]{arXiv:#2}

\bibitem[Adhikari and Sur(2024)]%
        {adhikari2024misleep}
\bibfield{author}{\bibinfo{person}{Aakriti Adhikari} {and} \bibinfo{person}{Sanjib Sur}.} \bibinfo{year}{2024}\natexlab{}.
\newblock \showarticletitle{MiSleep: Human sleep posture identification from deep learning augmented millimeter-wave wireless systems}.
\newblock \bibinfo{journal}{\emph{ACM Transactions on Internet of Things}} \bibinfo{volume}{5}, \bibinfo{number}{2} (\bibinfo{year}{2024}), \bibinfo{pages}{1--33}.
\newblock


\bibitem[Almazrouei et~al\mbox{.}(2023)]%
        {almazrouei2023falcon}
\bibfield{author}{\bibinfo{person}{Ebtesam Almazrouei}, \bibinfo{person}{Hamza Alobeidli}, \bibinfo{person}{Abdulaziz Alshamsi}, \bibinfo{person}{Alessandro Cappelli}, \bibinfo{person}{Ruxandra Cojocaru}, \bibinfo{person}{M{\'e}rouane Debbah}, \bibinfo{person}{{\'E}tienne Goffinet}, \bibinfo{person}{Daniel Hesslow}, \bibinfo{person}{Julien Launay}, \bibinfo{person}{Quentin Malartic}, {et~al\mbox{.}}} \bibinfo{year}{2023}\natexlab{}.
\newblock \showarticletitle{The falcon series of open language models}.
\newblock \bibinfo{journal}{\emph{arXiv preprint arXiv:2311.16867}} (\bibinfo{year}{2023}).
\newblock


\bibitem[Arnal et~al\mbox{.}(2020)]%
        {arnal2020dreem}
\bibfield{author}{\bibinfo{person}{Pierrick~J Arnal}, \bibinfo{person}{Valentin Thorey}, \bibinfo{person}{Eden Debellemaniere}, \bibinfo{person}{Michael~E Ballard}, \bibinfo{person}{Amadeu Bou~Hernandez}, \bibinfo{person}{Arthur Guillot}, \bibinfo{person}{Hugo Jourde}, \bibinfo{person}{Milton Harris}, \bibinfo{person}{Mathias Guillard}, \bibinfo{person}{Mathilde Van~Egroo}, {et~al\mbox{.}}} \bibinfo{year}{2020}\natexlab{}.
\newblock \showarticletitle{The Dreem Headband as an Alternative to Polysomnography for EEG Signal Acquisition and Sleep Staging}.
\newblock \bibinfo{journal}{\emph{Sleep}} \bibinfo{volume}{43}, \bibinfo{number}{11} (\bibinfo{year}{2020}).
\newblock


\bibitem[Bujan et~al\mbox{.}(2023)]%
        {bujan2023clinical}
\bibfield{author}{\bibinfo{person}{Bartosz Bujan}, \bibinfo{person}{Tobit Fischer}, \bibinfo{person}{Sarah Dietz-Terjung}, \bibinfo{person}{Aribert Bauerfeind}, \bibinfo{person}{Piotr Jedrysiak}, \bibinfo{person}{Martina Gro{\ss}e~Sundrup}, \bibinfo{person}{Janne Hamann}, {and} \bibinfo{person}{Christoph Sch{\"o}bel}.} \bibinfo{year}{2023}\natexlab{}.
\newblock \showarticletitle{Clinical validation of a contactless respiration rate monitor}.
\newblock \bibinfo{journal}{\emph{Scientific Reports}} \bibinfo{volume}{13}, \bibinfo{number}{1} (\bibinfo{year}{2023}), \bibinfo{pages}{3480}.
\newblock


\bibitem[De~Zambotti et~al\mbox{.}(2024)]%
        {de2024state}
\bibfield{author}{\bibinfo{person}{Massimiliano De~Zambotti}, \bibinfo{person}{Cathy Goldstein}, \bibinfo{person}{Jesse Cook}, \bibinfo{person}{Luca Menghini}, \bibinfo{person}{Marco Altini}, \bibinfo{person}{Philip Cheng}, {and} \bibinfo{person}{Rebecca Robillard}.} \bibinfo{year}{2024}\natexlab{}.
\newblock \showarticletitle{State of the science and recommendations for using wearable technology in sleep and circadian research}.
\newblock \bibinfo{journal}{\emph{Sleep}} \bibinfo{volume}{47}, \bibinfo{number}{4} (\bibinfo{year}{2024}).
\newblock


\bibitem[Deng et~al\mbox{.}(2018)]%
        {deng2018design}
\bibfield{author}{\bibinfo{person}{Fei Deng}, \bibinfo{person}{Jianwu Dong}, \bibinfo{person}{Xiangyu Wang}, \bibinfo{person}{Ying Fang}, \bibinfo{person}{Yu Liu}, \bibinfo{person}{Zhaofei Yu}, \bibinfo{person}{Jing Liu}, {and} \bibinfo{person}{Feng Chen}.} \bibinfo{year}{2018}\natexlab{}.
\newblock \showarticletitle{Design and implementation of a noncontact sleep monitoring system using infrared cameras and motion sensor}.
\newblock \bibinfo{journal}{\emph{IEEE Transactions on Instrumentation and Measurement}} \bibinfo{volume}{67}, \bibinfo{number}{7} (\bibinfo{year}{2018}), \bibinfo{pages}{1555--1563}.
\newblock


\bibitem[Fan et~al\mbox{.}(2019)]%
        {fan2019light}
\bibfield{author}{\bibinfo{person}{Junliang Fan}, \bibinfo{person}{Xin Ma}, \bibinfo{person}{Lifeng Wu}, \bibinfo{person}{Fucang Zhang}, \bibinfo{person}{Xiang Yu}, {and} \bibinfo{person}{Wenzhi Zeng}.} \bibinfo{year}{2019}\natexlab{}.
\newblock \showarticletitle{Light Gradient Boosting Machine: An efficient soft computing model for estimating daily reference evapotranspiration with local and external meteorological data}.
\newblock \bibinfo{journal}{\emph{Agricultural water management}}  \bibinfo{volume}{225} (\bibinfo{year}{2019}), \bibinfo{pages}{105758}.
\newblock


\bibitem[Han et~al\mbox{.}(2024)]%
        {han2024earsleep}
\bibfield{author}{\bibinfo{person}{Feiyu Han}, \bibinfo{person}{Panlong Yang}, \bibinfo{person}{Yuanhao Feng}, \bibinfo{person}{Weiwei Jiang}, \bibinfo{person}{Youwei Zhang}, {and} \bibinfo{person}{Xiang-Yang Li}.} \bibinfo{year}{2024}\natexlab{}.
\newblock \showarticletitle{Earsleep: In-ear acoustic-based physical and physiological activity recognition for sleep stage detection}.
\newblock \bibinfo{journal}{\emph{Proceedings of the ACM on Interactive, Mobile, Wearable and Ubiquitous Technologies}} \bibinfo{volume}{8}, \bibinfo{number}{2} (\bibinfo{year}{2024}), \bibinfo{pages}{1--31}.
\newblock


\bibitem[He et~al\mbox{.}(2024)]%
        {he2024novel}
\bibfield{author}{\bibinfo{person}{Chunhua He}, \bibinfo{person}{Shuibin Liu}, \bibinfo{person}{Zewen Fang}, \bibinfo{person}{Heng Wu}, \bibinfo{person}{Maojin Liang}, \bibinfo{person}{Songqing Deng}, {and} \bibinfo{person}{Juze Lin}.} \bibinfo{year}{2024}\natexlab{}.
\newblock \showarticletitle{A Novel Detection Method for Heart Rate Variability and Sleep Posture Based on a Flexible Sleep Monitoring Belt}.
\newblock \bibinfo{journal}{\emph{IEEE Sensors Journal}} (\bibinfo{year}{2024}).
\newblock


\bibitem[Hong et~al\mbox{.}(2019)]%
        {hong2019microwave}
\bibfield{author}{\bibinfo{person}{Hong Hong}, \bibinfo{person}{Li Zhang}, \bibinfo{person}{Heng Zhao}, \bibinfo{person}{Hui Chu}, \bibinfo{person}{Chen Gu}, \bibinfo{person}{Michael Brown}, \bibinfo{person}{Xiaohua Zhu}, {and} \bibinfo{person}{Changzhi Li}.} \bibinfo{year}{2019}\natexlab{}.
\newblock \showarticletitle{Microwave sensing and sleep: Noncontact sleep-monitoring technology with microwave biomedical radar}.
\newblock \bibinfo{journal}{\emph{IEEE Microwave Magazine}} \bibinfo{volume}{20}, \bibinfo{number}{8} (\bibinfo{year}{2019}), \bibinfo{pages}{18--29}.
\newblock


\bibitem[Islam and Lubecke(2022)]%
        {islam2022sleep}
\bibfield{author}{\bibinfo{person}{Shekh Md~Mahmudul Islam} {and} \bibinfo{person}{Victor~M Lubecke}.} \bibinfo{year}{2022}\natexlab{}.
\newblock \showarticletitle{Sleep posture recognition with a dual-frequency microwave Doppler radar and machine learning classifiers}.
\newblock \bibinfo{journal}{\emph{IEEE Sensors Letters}} \bibinfo{volume}{6}, \bibinfo{number}{3} (\bibinfo{year}{2022}), \bibinfo{pages}{1--4}.
\newblock


\bibitem[Jayarajah et~al\mbox{.}(2015)]%
        {jayarajah2015candy}
\bibfield{author}{\bibinfo{person}{Kasthuri Jayarajah}, \bibinfo{person}{Meera Radhakrishnan}, \bibinfo{person}{Steven Hoi}, {and} \bibinfo{person}{Archan Misra}.} \bibinfo{year}{2015}\natexlab{}.
\newblock \showarticletitle{Candy crushing your sleep}. In \bibinfo{booktitle}{\emph{Adjunct proceedings of the 2015 ACM international joint conference on pervasive and ubiquitous computing and proceedings of the 2015 ACM international symposium on wearable computers}}. \bibinfo{pages}{753--762}.
\newblock


\bibitem[Jiang et~al\mbox{.}(2023)]%
        {jiang2023mistral7b}
\bibfield{author}{\bibinfo{person}{Albert~Q. Jiang}, \bibinfo{person}{Alexandre Sablayrolles}, \bibinfo{person}{Arthur Mensch}, \bibinfo{person}{Chris Bamford}, \bibinfo{person}{Devendra~Singh Chaplot}, \bibinfo{person}{Diego de~las Casas}, \bibinfo{person}{Florian Bressand}, \bibinfo{person}{Gianna Lengyel}, \bibinfo{person}{Guillaume Lample}, \bibinfo{person}{Lucile Saulnier}, \bibinfo{person}{Lélio~Renard Lavaud}, \bibinfo{person}{Marie-Anne Lachaux}, \bibinfo{person}{Pierre Stock}, \bibinfo{person}{Teven~Le Scao}, \bibinfo{person}{Thibaut Lavril}, \bibinfo{person}{Thomas Wang}, \bibinfo{person}{Timothée Lacroix}, {and} \bibinfo{person}{William~El Sayed}.} \bibinfo{year}{2023}\natexlab{}.
\newblock \bibinfo{title}{Mistral 7B}.
\newblock
\showeprint[arxiv]{2310.06825}~[cs.CL]
\urldef\tempurl%
\url{https://arxiv.org/abs/2310.06825}
\showURL{%
\tempurl}


\bibitem[Karlgren et~al\mbox{.}(2022)]%
        {karlgren2022self}
\bibfield{author}{\bibinfo{person}{Kasper Karlgren}, \bibinfo{person}{Barry Brown}, {and} \bibinfo{person}{Donald McMillan}.} \bibinfo{year}{2022}\natexlab{}.
\newblock \showarticletitle{From self-tracking to sleep-hacking: online collaboration on changing sleep}.
\newblock \bibinfo{journal}{\emph{Proceedings of the ACM on human-computer interaction}} \bibinfo{volume}{6}, \bibinfo{number}{CSCW2} (\bibinfo{year}{2022}), \bibinfo{pages}{1--26}.
\newblock


\bibitem[Li et~al\mbox{.}(2025)]%
        {li2025novel}
\bibfield{author}{\bibinfo{person}{Denghao Li}, \bibinfo{person}{Yukun Huang}, \bibinfo{person}{Huaqing Li}, \bibinfo{person}{Jingran Cheng}, \bibinfo{person}{Wenwen Zhu}, {and} \bibinfo{person}{Haoming Feng}.} \bibinfo{year}{2025}\natexlab{}.
\newblock \showarticletitle{A Novel Approach to Accurate Respiratory Rate and Heart Rate Estimation via FMCW Radar}.
\newblock \bibinfo{journal}{\emph{IEEE Sensors Journal}} (\bibinfo{year}{2025}).
\newblock


\bibitem[Lin et~al\mbox{.}(2017)]%
        {lin2017iot}
\bibfield{author}{\bibinfo{person}{Chin-Teng Lin}, \bibinfo{person}{Mukesh Prasad}, \bibinfo{person}{Chia-Hsin Chung}, \bibinfo{person}{Deepak Puthal}, \bibinfo{person}{Hesham El-Sayed}, \bibinfo{person}{Sharmi Sankar}, \bibinfo{person}{Yu-Kai Wang}, \bibinfo{person}{Jagendra Singh}, {and} \bibinfo{person}{Arun~Kumar Sangaiah}.} \bibinfo{year}{2017}\natexlab{}.
\newblock \showarticletitle{IoT-based wireless polysomnography intelligent system for sleep monitoring}.
\newblock \bibinfo{journal}{\emph{Ieee Access}}  \bibinfo{volume}{6} (\bibinfo{year}{2017}), \bibinfo{pages}{405--414}.
\newblock


\bibitem[Lin et~al\mbox{.}(2016)]%
        {lin2016sleepsense}
\bibfield{author}{\bibinfo{person}{Feng Lin}, \bibinfo{person}{Yan Zhuang}, \bibinfo{person}{Chen Song}, \bibinfo{person}{Aosen Wang}, \bibinfo{person}{Yiran Li}, \bibinfo{person}{Changzhan Gu}, \bibinfo{person}{Changzhi Li}, {and} \bibinfo{person}{Wenyao Xu}.} \bibinfo{year}{2016}\natexlab{}.
\newblock \showarticletitle{SleepSense: A noncontact and cost-effective sleep monitoring system}.
\newblock \bibinfo{journal}{\emph{IEEE transactions on biomedical circuits and systems}} \bibinfo{volume}{11}, \bibinfo{number}{1} (\bibinfo{year}{2016}), \bibinfo{pages}{189--202}.
\newblock


\bibitem[Liu et~al\mbox{.}(2023b)]%
        {liu2023fmcw}
\bibfield{author}{\bibinfo{person}{Guoxiang Liu}, \bibinfo{person}{Xingguang Li}, \bibinfo{person}{Chunsheng Xu}, \bibinfo{person}{Lei Ma}, {and} \bibinfo{person}{Hongye Li}.} \bibinfo{year}{2023}\natexlab{b}.
\newblock \showarticletitle{FMCW radar-based human sitting posture detection}.
\newblock \bibinfo{journal}{\emph{IEEE Access}}  \bibinfo{volume}{11} (\bibinfo{year}{2023}), \bibinfo{pages}{102746--102756}.
\newblock


\bibitem[Liu et~al\mbox{.}(2023a)]%
        {liu2023posmonitor}
\bibfield{author}{\bibinfo{person}{Xiulong Liu}, \bibinfo{person}{Wei Jiang}, \bibinfo{person}{Sheng Chen}, \bibinfo{person}{Xin Xie}, \bibinfo{person}{Hankai Liu}, \bibinfo{person}{Qixuan Cai}, \bibinfo{person}{Xinyu Tong}, \bibinfo{person}{Tuo Shi}, {and} \bibinfo{person}{Wenyu Qu}.} \bibinfo{year}{2023}\natexlab{a}.
\newblock \showarticletitle{PosMonitor: Fine-grained sleep posture recognition with mmWave radar}.
\newblock \bibinfo{journal}{\emph{IEEE Internet of Things Journal}} \bibinfo{volume}{11}, \bibinfo{number}{7} (\bibinfo{year}{2023}), \bibinfo{pages}{11175--11189}.
\newblock


\bibitem[Mihalcea and Tarau(2004)]%
        {mihalcea2004textrank}
\bibfield{author}{\bibinfo{person}{Rada Mihalcea} {and} \bibinfo{person}{Paul Tarau}.} \bibinfo{year}{2004}\natexlab{}.
\newblock \showarticletitle{Textrank: Bringing order into text}. In \bibinfo{booktitle}{\emph{Proceedings of the 2004 conference on empirical methods in natural language processing}}. \bibinfo{pages}{404--411}.
\newblock


\bibitem[Nouman et~al\mbox{.}(2021)]%
        {nouman2021recent}
\bibfield{author}{\bibinfo{person}{Muhammad Nouman}, \bibinfo{person}{Sui~Yang Khoo}, \bibinfo{person}{MA~Parvez Mahmud}, {and} \bibinfo{person}{Abbas~Z Kouzani}.} \bibinfo{year}{2021}\natexlab{}.
\newblock \showarticletitle{Recent advances in contactless sensing technologies for mental health monitoring}.
\newblock \bibinfo{journal}{\emph{IEEE Internet of Things Journal}} \bibinfo{volume}{9}, \bibinfo{number}{1} (\bibinfo{year}{2021}), \bibinfo{pages}{274--297}.
\newblock


\bibitem[OpenAI(2023)]%
        {openai2023gpt4}
\bibfield{author}{\bibinfo{person}{OpenAI}.} \bibinfo{year}{2023}\natexlab{}.
\newblock \bibinfo{title}{GPT-4 Technical Report}.
\newblock \bibinfo{howpublished}{\url{https://openai.com/research/gpt-4}}.
\newblock
\newblock
\shownote{Accessed: 2025-05-23}.


\bibitem[Pandala(2025)]%
        {lazypredict}
\bibfield{author}{\bibinfo{person}{Shankar~Rao Pandala}.} \bibinfo{year}{2025}\natexlab{}.
\newblock \bibinfo{title}{Lazy Predict}.
\newblock \bibinfo{howpublished}{\href{https://lazypredict.readthedocs.io/en/latest/}{https://lazypredict.readthedocs.io/en/latest/}}.
\newblock
\newblock
\shownote{Accessed: 2025-05-23}.


\bibitem[Perez-Pozuelo et~al\mbox{.}(2020)]%
        {perez2020future}
\bibfield{author}{\bibinfo{person}{Ignacio Perez-Pozuelo}, \bibinfo{person}{Bing Zhai}, \bibinfo{person}{Joao Palotti}, \bibinfo{person}{Raghvendra Mall}, \bibinfo{person}{Micha{\"e}l Aupetit}, \bibinfo{person}{Juan~M Garcia-Gomez}, \bibinfo{person}{Shahrad Taheri}, \bibinfo{person}{Yu Guan}, {and} \bibinfo{person}{Luis Fernandez-Luque}.} \bibinfo{year}{2020}\natexlab{}.
\newblock \showarticletitle{The future of sleep health: a data-driven revolution in sleep science and medicine}.
\newblock \bibinfo{journal}{\emph{NPJ digital medicine}} \bibinfo{volume}{3}, \bibinfo{number}{1} (\bibinfo{year}{2020}), \bibinfo{pages}{42}.
\newblock


\bibitem[Piriyajitakonkij et~al\mbox{.}(2020)]%
        {piriyajitakonkij2020sleepposenet}
\bibfield{author}{\bibinfo{person}{Maytus Piriyajitakonkij}, \bibinfo{person}{Patchanon Warin}, \bibinfo{person}{Payongkit Lakhan}, \bibinfo{person}{Pitshaporn Leelaarporn}, \bibinfo{person}{Nakorn Kumchaiseemak}, \bibinfo{person}{Supasorn Suwajanakorn}, \bibinfo{person}{Theerasarn Pianpanit}, \bibinfo{person}{Nattee Niparnan}, \bibinfo{person}{Subhas~Chandra Mukhopadhyay}, {and} \bibinfo{person}{Theerawit Wilaiprasitporn}.} \bibinfo{year}{2020}\natexlab{}.
\newblock \showarticletitle{SleepPoseNet: Multi-view learning for sleep postural transition recognition using UWB}.
\newblock \bibinfo{journal}{\emph{IEEE Journal of Biomedical and Health Informatics}} \bibinfo{volume}{25}, \bibinfo{number}{4} (\bibinfo{year}{2020}), \bibinfo{pages}{1305--1314}.
\newblock


\bibitem[Touvron et~al\mbox{.}(2023)]%
        {touvron2023llama}
\bibfield{author}{\bibinfo{person}{Hugo Touvron}, \bibinfo{person}{Louis Martin}, \bibinfo{person}{Kevin Stone}, \bibinfo{person}{Peter Albert}, \bibinfo{person}{Amjad Almahairi}, \bibinfo{person}{Yasmine Babaei}, \bibinfo{person}{Nikolay Bashlykov}, \bibinfo{person}{Soumya Batra}, \bibinfo{person}{Prajjwal Bhargava}, \bibinfo{person}{Shruti Bhosale}, {et~al\mbox{.}}} \bibinfo{year}{2023}\natexlab{}.
\newblock \showarticletitle{Llama 2: Open foundation and fine-tuned chat models}.
\newblock \bibinfo{journal}{\emph{arXiv preprint arXiv:2307.09288}} (\bibinfo{year}{2023}).
\newblock


\bibitem[Wang et~al\mbox{.}(2024)]%
        {wang2024video}
\bibfield{author}{\bibinfo{person}{Qiongyan Wang}, \bibinfo{person}{Hanrong Cheng}, {and} \bibinfo{person}{Wenjin Wang}.} \bibinfo{year}{2024}\natexlab{}.
\newblock \showarticletitle{Video-psg: An intelligent contactless monitoring system for sleep staging}.
\newblock \bibinfo{journal}{\emph{IEEE Transactions on Biomedical Engineering}} (\bibinfo{year}{2024}).
\newblock


\bibitem[Wang and Shao(2022)]%
        {wang2022human}
\bibfield{author}{\bibinfo{person}{Xuyu Wang} {and} \bibinfo{person}{Dangdang Shao}.} \bibinfo{year}{2022}\natexlab{}.
\newblock \showarticletitle{Human physiology and contactless vital signs monitoring using camera and wireless signals}.
\newblock In \bibinfo{booktitle}{\emph{Contactless Vital Signs Monitoring}}. \bibinfo{publisher}{Elsevier}, \bibinfo{pages}{1--24}.
\newblock


\bibitem[Yao et~al\mbox{.}(2025)]%
        {yao2025mm2sleep}
\bibfield{author}{\bibinfo{person}{Yicheng Yao}, \bibinfo{person}{Hao Zhang}, \bibinfo{person}{Pan Xia}, \bibinfo{person}{Changyu Liu}, \bibinfo{person}{Fanglin Geng}, \bibinfo{person}{Zhongrui Bai}, \bibinfo{person}{Lidong Du}, \bibinfo{person}{Xianxiang Chen}, \bibinfo{person}{Peng Wang}, \bibinfo{person}{Weifeng Yao}, {et~al\mbox{.}}} \bibinfo{year}{2025}\natexlab{}.
\newblock \showarticletitle{mm2Sleep: Highly generalized dual-person sleep posture recognition using FMCW radar}.
\newblock \bibinfo{journal}{\emph{Biomedical Signal Processing and Control}}  \bibinfo{volume}{103} (\bibinfo{year}{2025}), \bibinfo{pages}{107430}.
\newblock


\bibitem[Yue et~al\mbox{.}(2020)]%
        {yue2020bodycompass}
\bibfield{author}{\bibinfo{person}{Shichao Yue}, \bibinfo{person}{Yuzhe Yang}, \bibinfo{person}{Hao Wang}, \bibinfo{person}{Hariharan Rahul}, {and} \bibinfo{person}{Dina Katabi}.} \bibinfo{year}{2020}\natexlab{}.
\newblock \showarticletitle{BodyCompass: Monitoring sleep posture with wireless signals}.
\newblock \bibinfo{journal}{\emph{Proceedings of the ACM on Interactive, Mobile, Wearable and Ubiquitous Technologies}} \bibinfo{volume}{4}, \bibinfo{number}{2} (\bibinfo{year}{2020}), \bibinfo{pages}{1--25}.
\newblock


\bibitem[Zhang et~al\mbox{.}(2017)]%
        {zhang2017learning}
\bibfield{author}{\bibinfo{person}{Shichao Zhang}, \bibinfo{person}{Xuelong Li}, \bibinfo{person}{Ming Zong}, \bibinfo{person}{Xiaofeng Zhu}, {and} \bibinfo{person}{Debo Cheng}.} \bibinfo{year}{2017}\natexlab{}.
\newblock \showarticletitle{Learning k for knn classification}.
\newblock \bibinfo{journal}{\emph{ACM Transactions on Intelligent Systems and Technology (TIST)}} \bibinfo{volume}{8}, \bibinfo{number}{3} (\bibinfo{year}{2017}), \bibinfo{pages}{1--19}.
\newblock


\end{thebibliography}

\end{document}